\documentclass[prd,preprintnumbers,superscriptaddress,nofootinbib,showpacs,twocolumn]{revtex4-2}
\usepackage{pstricks}
\usepackage{color}
\usepackage{amssymb,amsmath,bm,bbm}
\usepackage{epsf,epsfig}
\usepackage{afterpage}
\usepackage{longtable}
\usepackage{latexsym,mathrsfs,dsfont}
\usepackage{graphics}
\usepackage{url}
\usepackage{paralist}
\usepackage{bbold}
\usepackage{appendix}

\newcommand{\sm}{\textsc{SM}}
\newcommand{\np}{\textsc{NP}}
\newcommand{\inter}{\textsc{INT}}

\newcommand{\be}{\begin{equation}}
\newcommand{\ee}{\end{equation}}
\newcommand{\bi}{\begin{itemize}}
\newcommand{\ei}{\end{itemize}}
\newcommand{\ba}{\begin{array}}
\newcommand{\ea}{\end{array}}
\newcommand{\bea}{\begin{eqnarray}}
\newcommand{\eea}{\end{eqnarray}}
\newcommand{\bec}{\begin{center}}
\newcommand{\eec}{\end{center}}

\newcommand{\nn}{\nonumber}
\newcommand{\qq}{\quad \quad}

\def\@seccntformat#1{\@ifundefined{#1@cntformat}%
   {\csname the#1\endcsname\quad}
   {\csname #1@cntformat\endcsname}
}
\begin{document}

\preprint{BARI-TH/754-24}

\title{  New physics couplings  from angular coefficient functions of  $\bar B \to D^* (D \pi)  \ell \bar \nu_\ell$}
\author{Pietro~Colangelo}
\email[Electronic address:]{pietro.colangelo@ba.infn.it} 
\affiliation{Istituto Nazionale di Fisica Nucleare, Sezione di Bari, via Orabona 4, 70126 Bari, Italy}
\author{Fulvia~De~Fazio}
\email[Electronic address:]{fulvia.defazio@ba.infn.it} 
\affiliation{Istituto Nazionale di Fisica Nucleare, Sezione di Bari, via Orabona 4, 70126 Bari, Italy}
\author{Francesco~Loparco}
\email[Electronic address:]{francesco.loparco1@ba.infn.it} 
\affiliation{Istituto Nazionale di Fisica Nucleare, Sezione di Bari, via Orabona 4, 70126 Bari, Italy}
\author{Nicola~Losacco}
\email[Electronic address:]{nicola.losacco@ba.infn.it} 
\affiliation{Istituto Nazionale di Fisica Nucleare, Sezione di Bari, via Orabona 4, 70126 Bari, Italy}
\affiliation{Dipartimento Interateneo di Fisica "Michelangelo Merlin", Universit\`a degli Studi di Bari, via Orabona 4, 70126 Bari, Italy}

\begin{abstract}
\noindent
The Belle Collaboration has recently measured the complete set of angular coefficient functions for the exclusive decays $\bar B \to D^*  (D \pi)  \ell \bar \nu_\ell$, with $\ell=e,\,\mu$, in four bins of the parameter $w=\displaystyle\frac{m_B^2+m_{D^*}^2-q^2}{2m_B m_{D^*}}$, with $q$   the lepton pair momentum \cite{Belle:2023xgj}.
Under the assumption that  physics  beyond the Standard Model   does not contribute to such modes,  the measurements are useful to determine the hadronic form factors describing the $B \to D^*$  matrix elements of the Standard Model  weak current, and to improve the determination of $|V_{cb}|$. 
On the other hand,  they can be used to assess the impact of   possible new physics contributions. In  a bottom-up approach, we extend  the Standard Model  effective Hamiltonian  governing this mode  with the inclusion of  the full set of  Lorentz invariant  d=6 operators  compatible with the  gauge symmetry of the theory. The measured angular coefficient functions can  tightly constrain the  couplings in the generalized Hamiltonian.  We present the first results of this analysis, discussing how  improvements can be achieved when more  complete data on the angular coefficient functions  will be available.
\end{abstract}

\thispagestyle{empty}


\maketitle

\section{ Introduction and framework}
 In the search of signals from physics beyond the Standard Model (SM), a few  tensions between SM expectations and measurements have emerged in the flavour sector  
\cite{HFLAV:2022pwe,DeFazio:2023lmy}.
In particular, in addition to processes  suppressed at tree-level in  SM,  which are highly sensitive to virtual contribution of  heavy quanta \cite{Buras:2020xsm},  also charged current processes are under scrutiny  after the emergence of  anomalies  in the ratios $R(D^{(*)})=\displaystyle\frac{{\cal B}(B \to D^{(*)} \tau \,\nu_\tau)}{{\cal B}(B \to D^{(*)} \ell \, \nu_\ell)}$ ($\ell=e,\,\mu$)  in BaBar Collaboration \cite{Lees:2012xj} and     subsequent analyses \cite{BaBar:2013mob,Belle:2015qfa,LHCb:2015gmp,Belle:2016dyj,LHCb:2017smo,LHCb:2017rln,Belle:2019rba} (see \cite{Gambino:2020jvv,HFLAV:2022pwe} for averages and review).
The possibility of relating such anomalies to the tensions for the different determinations of $|V_{cb}|$  makes the investigation of such processes even more intriguing \cite{Colangelo:2016ymy}. 

The angular coefficient functions in the  fully differential  $\bar B(p) \to D^{(*)}(p^\prime,\,\epsilon)(D \pi) \ell (k_1){\bar \nu}_\ell(k_2)$ decay distribution
 are suitable observables to look for the  effects of new physics (NP) \cite{Colangelo:2018cnj,Bhattacharya:2018kig,Murgui:2019czp,Becirevic:2019tpx,Bobeth:2021lya}. 
The Belle Collaboration has recently reported the measurement of the complete set of such functions in four bins of the hadronic recoil parameter $w=\displaystyle\frac{m_B^2+m_{D^*}^2-q^2}{2m_B m_{D^*}}$, with $q=p-p^\prime$ \cite{Belle:2023xgj}. Here, we want to consider the role of NP in this process  using theoretical expressions that can  be applied also to other  modes \cite{Colangelo:2019axi,Colangelo:2021dnv}.
 We make use of  the Standard Model effective field theory (SMEFT) as  a model-independent framework to analyze NP contributions to beauty hadron decays \cite{Buchmuller:1985jz,Grzadkowski:2010es}.  If the NP scale $\Lambda_{NP}$  is much larger than the electroweak scale, the new massive degrees of freedom can be integrated out providing  an effective  Hamiltonian   in terms of  SM fields,   invariant under the SM gauge group.
In the extended  Hamiltonian   new operators  not present in the SM appear,  suppressed by  powers of $1/\Lambda_{NP}$. At   ${\cal O}\left(1/\Lambda_{NP}^2 \right)$ these are  dimension-six  operators. Among these, those relevant for the present study are four fermion operators.

To describe the modes  ${\bar B} \to V \ell^-  {\bar \nu}_\ell$, with $V$ a  meson comprising an up-type quark $U$, we consider  the generalized effective Hamiltonian 
 \bea
&&H_{\rm eff}^{b \to U \ell \nu}= {G_F \over \sqrt{2}} V_{Ub}\nn \\ &\times &\Big\{(1+\epsilon_V^\ell) \left({\bar U} \gamma_\mu (1-\gamma_5) b \right)\left( {\bar \ell} \gamma^\mu (1-\gamma_5) {\nu}_\ell \right)
\nn \\
&+&\epsilon_R^\ell \left({\bar U} \gamma_\mu (1+\gamma_5) b \right)\left( {\bar \ell} \gamma^\mu (1-\gamma_5) {\nu}_\ell \right) \label{heff}  \\
&+&  \epsilon_S^\ell \, \left({\bar U}  b\right)  \left({\bar \ell} (1-\gamma_5) { \nu}_\ell \right) 
+ \epsilon_P^\ell \, \left({\bar U} \gamma_5 b\right)  \left({\bar \ell} (1-\gamma_5) { \nu}_\ell \right) \nn \\
&+& \epsilon_T^\ell \, \left({\bar U} \sigma_{\mu \nu} (1-\gamma_5) b\right) \,\left( {\bar \ell} \sigma^{\mu \nu} (1-\gamma_5) { \nu}_\ell \right) \Big\} + h.c.\,\,\, , \nn
\eea
with $G_F$  the Fermi constant and $V_{Ub}$ the relevant element of the Cabibbo-Kobayashi-Maskawa (CKM) matrix. Besides the SM  term,  the low energy Hamiltonian \eqref{heff} comprises NP operators with complex  $\epsilon^\ell_{V,R,S,P,T}$  lepton-flavour dependent  coefficients.   The scalar operator does not contribute if $V$ is a vector meson, as $D^*$ in the present case.
\begin{figure}[b]
\begin{center}
\includegraphics[width = 0.4\textwidth]{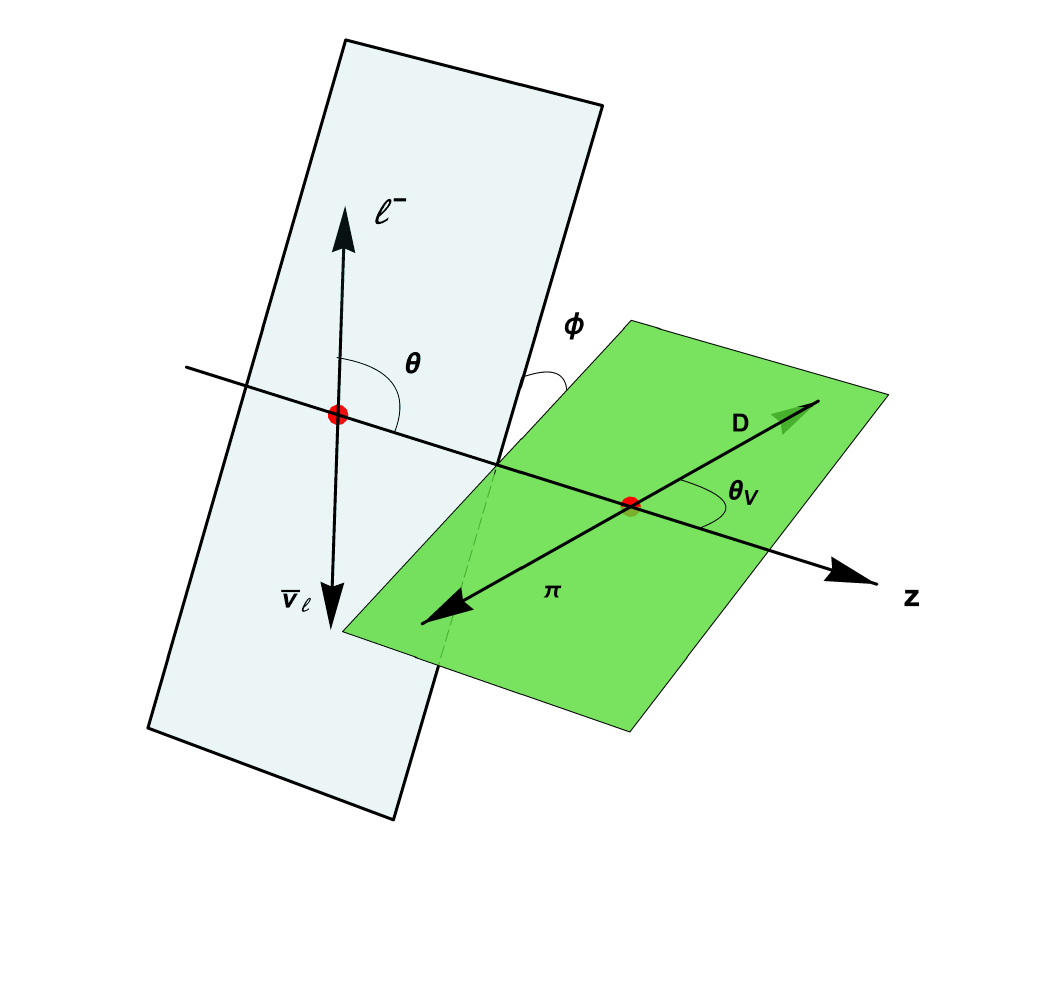}\vspace*{-1.cm}
\caption{\baselineskip 10pt  \small  Kinematics of  $\bar B \to D^*(D\pi) \ell^- \bar \nu_\ell$.}\label{fig:angles}
\end{center}
\end{figure}

We choose the kinematics  indicated in Fig.~\ref{fig:angles}: $\theta_V$ is the angle between the $D$ meson and the direction opposite to the ${\bar B}$ in the $D^*$  rest frame; $\theta_\ell$ the angle between the charged lepton and the ${\bar B}$ in the virtual $W$ rest frame and $\phi$ the angle between the decay planes identified by the directions of  the lepton pair on one hand and the $(D,\pi)$ pair on the other in the ${\bar B}$ rest frame. Using such variables the fully differential decay width reads:
\bea
&&\frac{d^4 \Gamma (\bar B \to V( P_1 P_2) \ell^- \bar \nu_\ell)}{dq^2 \,d\cos \theta \,d\phi \,d \cos \theta_V} 
={\cal C}|{\vec p}_{V}| \left(1-  \frac{ m_\ell^2}{q^2}\right)^2 \nn \\
&\times&\Big\{I_{1s} \,\sin^2 \theta_V+I_{1c} \,\cos^2\theta_V \nn \\
&+&\left(I_{2s} \,\sin^2 \theta_V+I_{2c} \,\cos^2 \theta_V\right) \cos 2\theta \nn  \\ 
&+&I_3 \,\sin^2 \theta_V \sin^2 \theta  \cos 2 \phi +I_4 \, \sin 2\theta_V \sin 2\theta \cos  \phi \nn \\  
&+&I_5 \, \sin 2 \theta_V \sin \theta \cos  \phi \label{angular} \\
&+&\left(I_{6s} \,\sin^2 \theta_V+I_{6c} \,\cos^2\theta_V\right)\cos \theta \nn \\
&+& I_7 \sin 2 \theta_V \sin \theta \sin  \phi+I_8 \,\sin 2 \theta_V \, \sin 2\theta \sin  \phi \nn \\
&+&I_9\,\sin^2 \theta_V \sin^2 \theta \sin  2\phi
  \Big\}\,\,, \nn
\eea
with
 ${\cal C}=\displaystyle{\frac{3G_F^2 |V_{Ub}|^2 {\cal B}(V \to P_1 P_2)}{128(2\pi)^4m_B^2}}$ and ${\vec p}_{V}$  the three-momentum of the $V$ meson (here $D^*$) in the $B$ meson rest-frame.
The angular coefficient functions $I_i$ in \eqref{angular} depend on the couplings $\epsilon^\ell_{V,R,P,T}$, on $q^2$ (or $w$) and on the hadronic form factors:
\bea
I_i &=& |1+\epsilon_V|^2 \,I_i^{SM}+|\epsilon_R|^2I_i^{NP,R}+|\epsilon_P|^2I_i^{NP,P} \hspace*{1cm} \nn \\
&+&|\epsilon_T|^2I_i^{NP,T} +2 \, {\rm Re}\left[\epsilon_R(1+\epsilon_V^* )\right] I_i^{INT,R} \nn \\
&+&2 \, {\rm Re}\left[\epsilon_P(1+\epsilon_V^* )\right] I_i^{INT,P} \nn \\
&+&2 \, {\rm Re}\left[\epsilon_T(1+\epsilon_V^* )\right] I_i^{INT,T} \label{eq:Iang1} \\
&+&2 \, {\rm Re}\left[\epsilon_R \epsilon_T^* \right] I_i^{INT,RT} 
+2 \, {\rm Re}\left[\epsilon_P \epsilon_T^* \right] I_i^{INT,PT}\nn \\
&+&2 \, {\rm Re}\left[\epsilon_P \epsilon_R^* \right] I_i^{INT,PR} \nn
\eea
for $ i=1s,\dots 6c$,
\bea
I_7 &=&2 \, {\rm Im}\left[\epsilon_R(1+\epsilon_V^* )\right] I_7^{INT,R} \nn \\
&+&2 \, {\rm Im}\left[\epsilon_P(1+\epsilon_V^* )\right] I_7^{INT,P}\nn \\
&+&2 \, {\rm Im}\left[\epsilon_T(1+\epsilon_V^* )\right] I_7^{INT,T} \label{eq:Iang2} \\
&+&2 \, {\rm Im}\left[\epsilon_R \epsilon_T^* \right] I_7^{INT,RT}+2 \, {\rm Im}\left[\epsilon_P \epsilon_T^* \right] I_7^{INT,PT} \qq \nn \\
&+&2 \, {\rm Im}\left[\epsilon_P \epsilon_R^* \right] I_7^{INT,PR}  , \nn
\eea
and for $ i=8,\,9$
\be
I_i=2 \, {\rm Im}\left[\epsilon_R (1+\epsilon_V^* ) \right] I_i^{INT,R} \,\,\, . 
\label{eq:Iang3} \ee
In SM such functions are expressed in terms of  helicity amplitudes:
\bea
H_0 &=&\frac{1}{{2m_V(m_B+m_V) \sqrt{q^2}}} \nn \\
&& \Big( (m_B+m_V)^2(m_B^2-m_V^2-q^2) A_1(q^2)\nn \\
&-&\lambda(m_B^2,\,m_V^2,\,q^2) A_2(q^2) \Big) \label{HampV}  \\
H_\pm&=& \frac{(m_B+m_V)^2 A_1(q^2)\mp\sqrt{\lambda(m_B^2,\,m_V^2,\,q^2)}V(q^2)}{m_B+m_V}  \nn  \\
H_t&=& -\frac{\sqrt{\lambda(m_B^2,\,m_V^2,\,q^2)}}{\sqrt{q^2}} \,A_0(q^2) , \,\,\,  \nn
\eea
with the form factors defined in the appendix \ref{app-ff}.
For  NP operators  the  amplitudes are also introduced:
\bea
&&H_\pm^{NP} =
 \frac{1}{\sqrt{q^2}}\Big\{q^2(T_1(q^2)- T_2(q^2))\nn \\
&+&\Big(m_B^2-m_V^2 \pm \sqrt{\lambda(m_B^2,m_V^2,q^2)} \Big)(T_1(q^2)+ T_2(q^2))\Big\} \nn \\
&&
H_L^{NP}= 4\Big\{
\frac{\lambda (m_B^2,m_V^2,q^2)}{m_V(m_B+m_V)^2} \, T_0(q^2)\nn \\
&+&2\frac{m_B^2+m_V^2-q^2}{m_V} T_1(q^2)+4 m_V T_2(q^2)\Big\}  . \label{HampNP}
\eea
The  form factors $T_i$ are also defined in appendix \ref{app-ff}. The  expressions of all coefficient functions $I_i$ are  in Tables I-V included the appendix \ref{app:coeff}.

\section{Constraints on  NP couplings from  Belle measurement}
\begin{figure}[ht]
\begin{center}
\includegraphics[width = 0.34\textwidth]{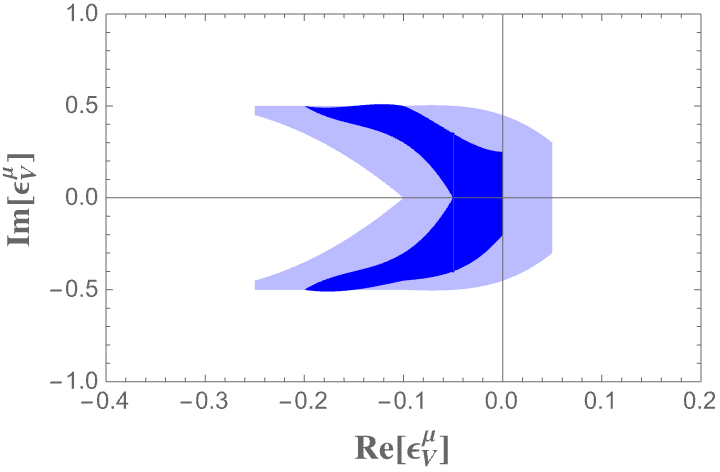}\\
\includegraphics[width = 0.34\textwidth]{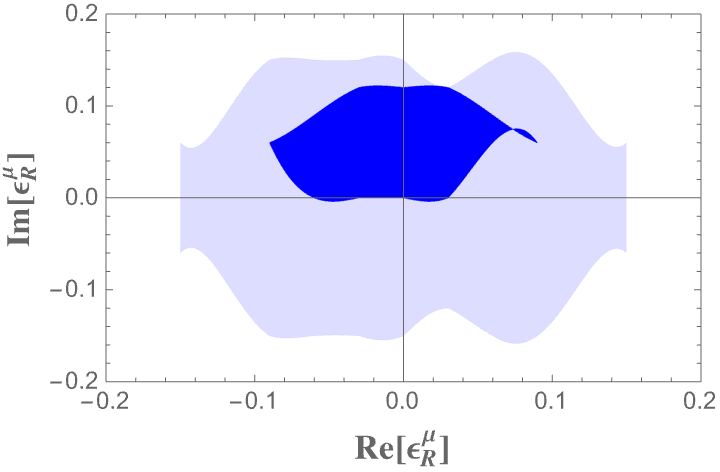}\\
\includegraphics[width = 0.34\textwidth]{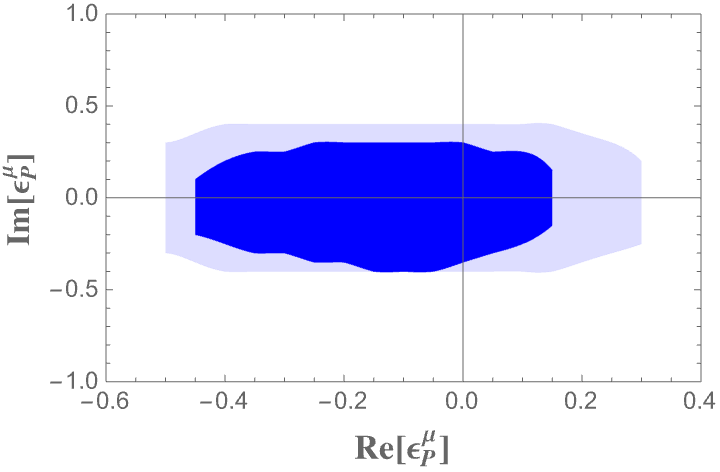}\\
\includegraphics[width = 0.34\textwidth]{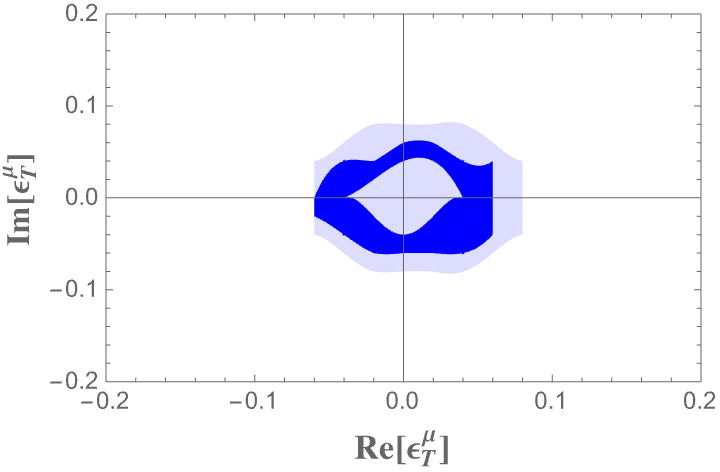}\\
\caption{\baselineskip 10pt  \small NP couplings  $\epsilon_V^\mu,\, \epsilon_R^\mu,\,\epsilon_P^\mu,\,\epsilon_T^\mu$ obtained from the Belle measurement of  the angular coefficient functions of  $\bar B \to D^*(D\pi) \mu^- \bar \nu_\mu$. The light region corresponds to requiring that the theoretical results agree with the experimental data at  2.5 $\sigma$. The dark region corresponds to the values obtained  minimizing the reduced $\chi^2$. }\label{fig:epsilon}
\end{center}
\end{figure}
 The comparison of  the Belle measurement \cite{Belle:2023xgj} to the theoretical expressions in the previous section allows us to constrain the couplings in the generalized Hamiltonian \eqref{heff}.
 Before presenting the analysis,  we point out that
i) the coefficient functions defined in the previous section are related to the coefficient functions in  \cite{Belle:2023xgj},  denoted as  $J_i$, through  $|I_i|=|J_i /F|$, with  $F=\displaystyle\frac{3 |{\vec p}_{D^*}|}{2^{ 10}m_B^5}$;  ii) the angle $\theta_\ell$ in \cite{Belle:2023xgj} corresponds  to $\theta_\ell \to \pi - \theta$ in our notation, therefore we have  $I_i=-J_i/F$ for $i=4,\,6s,\,6c,\,8$, and $I_i=J_i/F$ for all  the other  coefficients;
iii) the Belle Collaboration provides the angular coefficient functions in four bins $\Delta w^{(a)}$ of $w$, defining  ${\bar J}_i^a=\int_{{\Delta w}^{(a)}} J_i(w) dw$ and ${\hat J}_i=J_i/{\cal N}$, where ${\cal  N}=\frac{8}{9}\pi \sum_{a=1}^4 \left(3{\bar J}_{1c}^a+6{\bar J}_{1s}^a-{\bar J}_{2c}^a-2{\bar J}_{2s}^a \right)$. The factor ${\cal N}$ corresponds, modulo a constant, to the integrated width.

 To get  information on the effective couplings  we proceed as follows. From Fig.~1 of \cite{Belle:2023xgj}  we obtain the values of the coefficients ${\hat J}_i$ in the four bins ${\Delta w}^{(1)}=[1,\, 1.15]$, ${\Delta w}^{(2)}=[1.15,\, 1.25]$, ${\Delta w}^{(3)}=[1.25,\, 1.35]$, ${\Delta w}^{(4)}=[1.35,\, 1.5]$  (for such an information  the Collaboration has not provided in \cite{Belle:2023xgj} the table of numerical results and the error covariance matrix).
Next, we consider the products $({\hat J}_i^a )^{\rm exp}_{\rm int}={\hat J}_i \cdot (\Delta w)^a$.
Using the results  in appendix \ref{app:coeff} and fixing the parameter  ${\cal N}=0.146$ to reproduce the measured branching ratio quoted in \cite{Workman:2022ynf}, we calculate the expressions  of the integrals  of the coefficient functions in each bin of $w$, $({\hat J}_i^a)^{\rm th}_{\rm int}$. The $w$ dependence of the hadronic form factors is needed to perform the integrals: we use the CLN parametrization \cite{Caprini:1997mu} with the  parameters obtained  in \cite{Belle:2017rcc}.  The details on the reconstruction of the form factors and on the choice of the parameters can be found in  \cite{Colangelo:2018cnj}. 

We require that $({\hat J}_i^a)^{\rm th}_{\rm int} \in [({\hat J}_i^a )^{\rm exp}_{\rm int} - k\, \sigma_i^a,\, ({\hat J}_i^a )^{\rm exp}_{\rm int} +k\, \sigma_i^a]$, with $k$ a number of standard deviations to be fixed;  $\sigma_i^a$ is the error of the Belle result for each ${\hat J}_i^a$ multiplied by the corresponding bin width.  We have a total of 48 constraints, i.e. the  integrals over 4 bins for 12 angular coefficients. Since the data refer to the muon channel, we determine the set $(\epsilon_V^\mu,\, \epsilon_R^\mu,\,\epsilon_P^\mu,\,\epsilon_T^\mu$)  (set 1) that can simultaneously satisfy all  constraints, within the initial ranges  $|\epsilon_i^\mu|\le 0.5$ for $i=V,\,R,\,P\, ,T$. We find that the  smallest value of $k$ for which all  constraints are fulfilled is  $k=2.5$. 
For this set of parameters we compute the function $\chi^2_{red}=\displaystyle\frac{1}{N_{dof}}\sum_{i,\,a}{ \left( ({\hat J}_i^a)^{\rm th}_{\rm int}-({\hat J}_i^a)^{\rm exp}_{\rm int} \right) ^2}/{\left (\sigma_i^a \right)^2}$, with $N_{dof}=40$,  $i=1s,...9$ and $a=1,...4$.
The values of this function lie in the range $[1.8,2.6]$, the minimum corresponds to the set of parameters $\big({\rm Re}[\epsilon_V],{\rm Im}[\epsilon_V]\big)=\big(-0.05,0.25 \big)$; $\big({\rm Re}[\epsilon_R),{\rm Im}[\epsilon_R]\big)=\big(-0.03,0.09 \big)$;  $\big({\rm Re}[\epsilon_P],{\rm Im}[\epsilon_P]\big)=\big(-0.25,-0.25 \big)$; $\big({\rm Re}[\epsilon_T],{\rm Im}[\epsilon_T]\big)=\big(0.04,-0.04 \big)$. To be conservative
  we select more points,  those satisfying  $\chi^2_{red}\le  1.875$
(set 2). This choice is sensible since the probability to find $\chi^2_{red}>1.875$ in the case of 40  degrees of freedom is $0.07\%$. For comparison, the SM case (i.e.  $\epsilon_V=\epsilon_R=\epsilon_P=\epsilon_T=0$ and $N_{dof}=48$) corresponds to $\chi^2_{red}=2$.
The results are in Fig.~\ref{fig:epsilon}:  the light region corresponds to the parameters in set 1, while the dark region to the parameters in set 2.
%
The agreement with  data can be appreciated from Fig.~\ref{angcoeff} which includes the Belle points together with the angular coefficient functions obtained using the determined $\epsilon$ couplings. The remarkable result is that, while the SM point with all new couplings equal to zero is allowed,   it does not belong to the region of minimum $\chi^2$, and there is the possibility of values different from zero  in  $\epsilon_T^\mu$.
This observation will be strengthened when the table of measurements and the error covariance matrix will be available.

%
\begin{widetext}
\onecolumngrid
\begin{figure}[t!]
\begin{center}
\includegraphics[width = 0.88\textwidth]{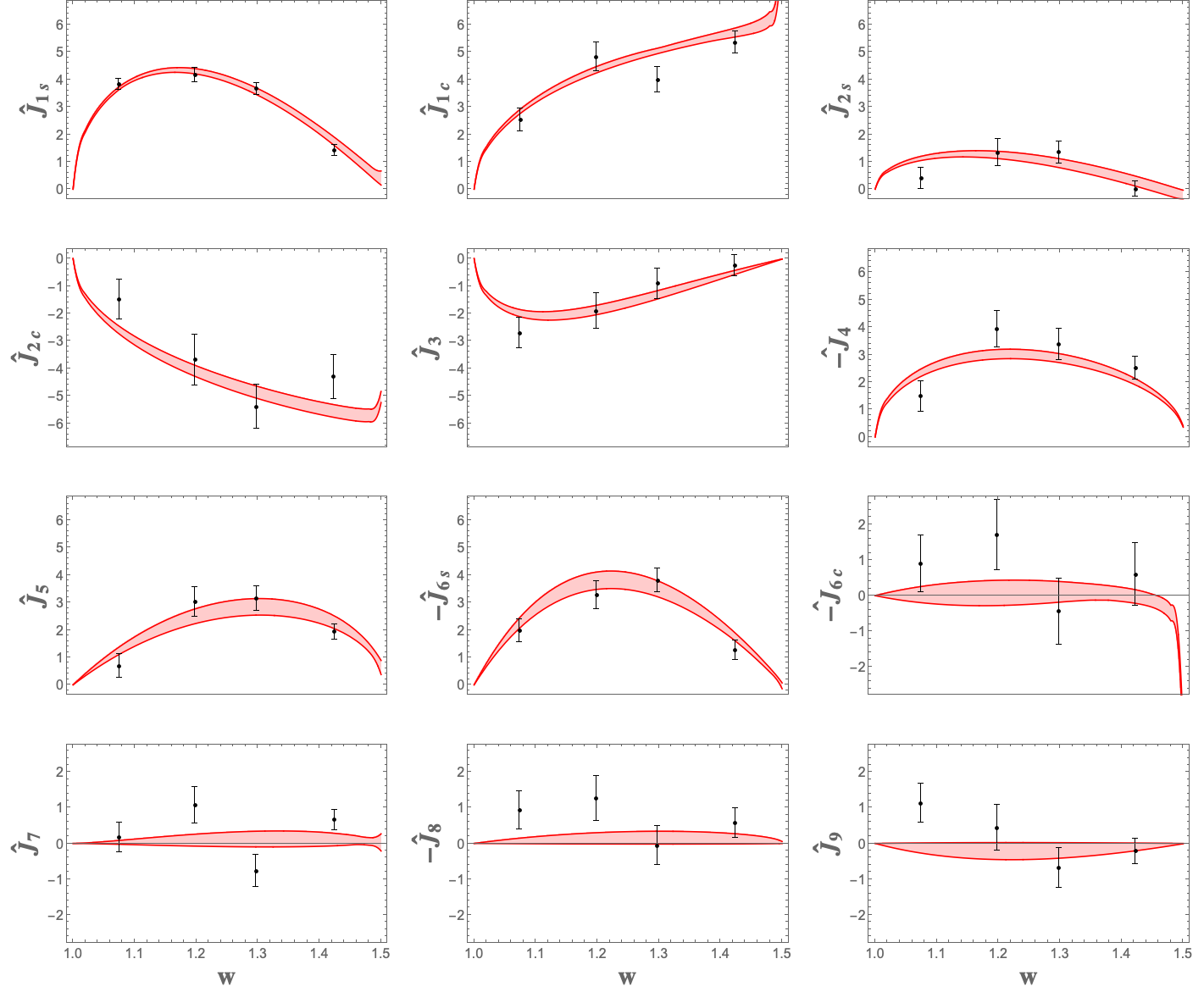}
\caption{\baselineskip 10pt  \small Angular coefficient functions in Eq.~\eqref{angular} for  $\bar B \to D^*(D\pi) \mu^- \bar \nu_\mu$. The shaded regions correspond to the results obtained using the determined  $\epsilon$ couplings, the points are the Belle measurements \cite{Belle:2023xgj}. }\label{angcoeff}
\end{center}
\end{figure}
\twocolumngrid
\end{widetext}

%

\begin{figure}[b]
\includegraphics[width = 0.4\textwidth]{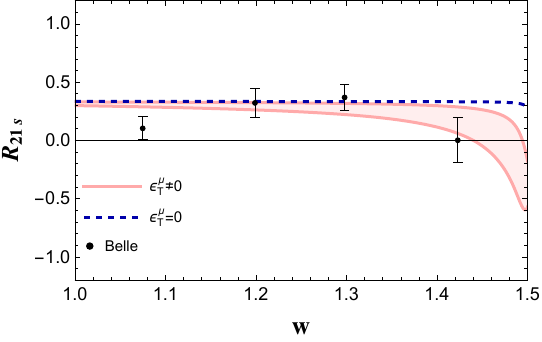}
\caption{\baselineskip 10pt  \small Ratio 
$R_{21s}$ in Eq.~\eqref{R21s}
for $\epsilon_T^\mu=0$ (dashed line) and  varying $\epsilon_T^\mu$ in the range displayed in Fig.~\ref{fig:epsilon} (shaded region). 
}\label{fig:R21s}
\end{figure}

Other observables  are sensitive to  the effects  of the new operators in \eqref{heff}. In particular, an interesting observable  is the ratio
\be
R_{21s}(w)=\frac{{\hat J}_{2s}(w)}{{\hat J}_{1s}(w)}\label {R21s} \,\,.
\ee
Indeed, as obtained using the expressions  in appendix \ref{app:coeff}, the angular coefficient functions $I_{1s,\,2s}$, hence  ${\hat J}_{1s,\,2s}$, do not depend on $\epsilon_P$. For vanishing $\epsilon_T$  their ratio would be independent of the form factors and insensitive to $\epsilon_V,\,\epsilon_R$. Therefore, the ratio \eqref {R21s} might signal  the tensor operator. This  is displayed in Fig.~\ref{fig:R21s} which shows that for nonvanishing $\epsilon_T$ the ratio can have a zero for $w_0$  in the range $[1.44,\,1.5]$  in the muon channel.

\begin{figure}[t]
\includegraphics[width = 0.4\textwidth]{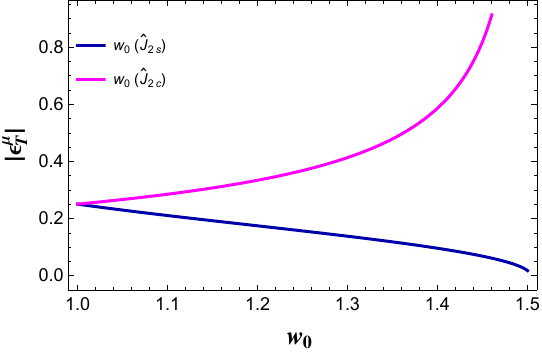}
\caption{\baselineskip 10pt  \small  $|\epsilon_T^\mu |$ as a function of the position of the zero of the ratio ${\hat J}_{2s}$ (blue curve) and  ${\hat J}_{2c}$ (magenta curve) for $\epsilon_V^\mu=\epsilon_R^\mu=0$.
}\label{fig:absepsT}
\end{figure}

The structure of the angular coefficients provides insights on the possibility that  
 $\epsilon_T$ is the only nonvanishing new coupling. In particular, for $\epsilon_V=\epsilon_R=0$ one finds that both  ${\hat J}_{2s}$  and  ${\hat J}_{2c}$  might have a zero. This is  depicted in Fig.~\ref{fig:absepsT}. This figure shows that if $|\epsilon_T^\mu|<0.25$ ${\hat J}_{2s}$ should have a zero while ${\hat J}_{2c}$ should not, and viceversa.  They cannot have a zero simultaneously. 
Although the Belle data  are not precise enough to draw definite conclusions,  they seem to exclude the presence of a zero in ${\hat J}_{2c}$ and are compatible with the presence of a zero of ${\hat J}_{2s}$ in the last bin of $w$.
\begin{figure}[b]
\begin{center}
\includegraphics[width = 0.4\textwidth]{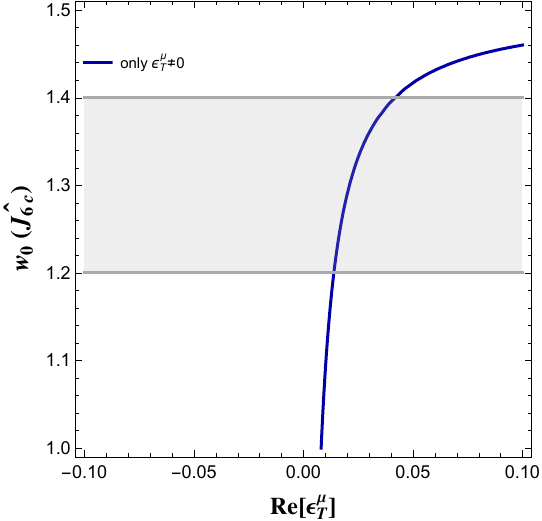}
\caption{\baselineskip 10pt  \small  $ {\rm Re}[\epsilon_T^\mu]$ as a function of the position of the zero of  ${\hat J}_{6c}$ for $\epsilon_V^\mu=\epsilon_R^\mu=\epsilon_P^\mu=0$, obtained from Eq.~(\ref{I6c}).
}\label{zeroI6c}
\end{center}
\end{figure}
Other angular coefficient functions provide us with further information. If the only nonvanishing NP coupling  is $\epsilon_T$, ${\hat J}_{6c}$ would display a zero  at a value given by the relation
\be
\sqrt{q^2}H_L^{\np}(q^2) {\rm Re}[\epsilon_T]-4m_\ell H_0(q^2)=0\,\,\, . \label{I6c}
\ee
The position $w_0$ of the zero of ${\hat J}_{6c}$ would fix ${\rm Re}[\epsilon_T]$. This is shown in Fig.~\ref{zeroI6c}, where the continuous  curve corresponds to the relation ~\eqref{I6c}, while the gray band is  the range of $w$ where the zero of ${\hat J}_{6c}$ should be found according to the Belle measurement. This corresponds to small values of ${\rm Re}[\epsilon_T^\mu]$, consistently with the results for ${\hat J}_{2s}$  and  ${\hat J}_{2c}$.

\section{Conclusions}
 The measurement of the full set of angular coefficient functions in $\bar B \to D^* (D \pi)  \mu \bar \nu_\mu$ constrains the set of NP coefficients in the generalized low energy Hamiltonian. In particular,  the possibility that some NP coefficients are different from zero emerges as a first evidence  on the basis of the information available in \cite{Belle:2023xgj}. It will be possible to corroborate this indication using
the experimental table of measurements together with the covariance error matrix.
\vspace*{0.5cm}

\noindent {\bf Note added.} When this manuscript  was completed, the paper \cite{Kapoor:2024ufg} was uploaded to the archive,  dealing with the same issue.

\vspace*{0.5cm}
\noindent {\bf Acknowledgements.}
We thank M. Prim and M. Rotondo for  discussions.
This study has been  carried out within the INFN project (Iniziativa Specifica) SPIF.
The research has been partly funded by the European Union – Next Generation EU through the research Grant No. P2022Z4P4B “SOPHYA - Sustainable Optimised PHYsics Algorithms: fundamental physics to build an advanced society" under the program PRIN 2022 PNRR of the Italian Ministero dell’Universit\'a e Ricerca (MUR).
\appendix
\numberwithin{equation}{section}
\section{Hadronic matrix elements}\label{app-ff}
The  $\bar B \to V$ matrix elements are parametrized as:
\bea
&&\langle V(p^\prime,\epsilon)|{\bar U} \gamma_\mu b| {B}(p) \rangle = 
- {2 V(q^2) \over m_{B}+m_V} i \epsilon_{\mu \nu \alpha \beta} \epsilon^{*\nu}  p^\alpha p^{\prime \beta}, \nn 
\eea
\bea
&&\langle V(p^\prime,\epsilon)|{\bar U} \gamma_\mu\gamma_5 b| {B}(p) \rangle = \nn \\
&&(m_{B}+m_V) \Big( \epsilon^*_\mu -{(\epsilon^* \cdot q) \over q^2} q_\mu \Big) A_1(q^2) \nn\\
&&- {(\epsilon^* \cdot q) \over  m_{B}+m_V} \Big( (p+p^\prime)_\mu -{m_{B}^2-m_V^2 \over q^2} q_\mu \Big) A_2(q^2) \nn \\
&&+ (\epsilon^* \cdot q){2 m_V \over q^2} q_\mu A_0(q^2),   \\
&&\langle V(p^\prime,\epsilon)|{\bar U} \gamma_5 b| {B}(p) \rangle =-\frac{2 m_V}{m_b+m_U} (\epsilon^* \cdot q) A_0(q^2), \nn
\eea
with the condition  
\bea  
A_0(0)&=&  \frac{m_{B} + m_V}{2 m_V} A_1(0) -  \frac{m_{B} - m_V}{2 m_V}  A_2(0) , \qq \qq
\eea
and

\bea
&&\langle V(p^\prime,\epsilon)|{\bar U} \sigma_{\mu \nu} b| { B}(p) \rangle =\nn \\
&& T_0(q^2) {\epsilon^* \cdot q \over (m_{B}+ m_V)^2} \epsilon_{\mu \nu \alpha \beta} p^\alpha p^{\prime \beta}\nn \\
&& +T_1(q^2) \epsilon_{\mu \nu \alpha \beta} p^\alpha \epsilon^{*\beta}
 + T_2(q^2) \epsilon_{\mu \nu \alpha \beta} p^{\prime \alpha} \epsilon^{*\beta}, 
\nn \\ 
&&\langle V(p^\prime,\epsilon)|{\bar U} \sigma_{\mu \nu}\gamma_5 b| { B}(p) \rangle =  \\
&&\qq i\, T_0(q^2) {\epsilon^* \cdot q \over (m_{B}+ m_V)^2} (p_\mu p^\prime_\nu-p_\nu p^\prime_\mu) \nn \\
&&\qq +i\,
T_1(q^2) (p_\mu \epsilon_\nu^*-\epsilon_\mu^* p_\nu) 
+i\,T_2(q^2)(p^\prime_\mu \epsilon_\nu^*-\epsilon_\mu^* p^\prime_\nu) . \nn
\eea

\begin{widetext}
\section{Angular coefficient functions}\label{app:coeff}

\begin{table}[h]
\caption{  \small Angular coefficient functions  in the 4d   $\bar B \to V (P_1 P_2)   \ell^- \bar \nu_\ell$ decay distribution  for the SM.}\label{tab:SM}
\vspace{0.3cm}
\centering
\begin{tabular}{cc}
\hline
\hline
\noalign{\medskip}
$i$ & $I_i^{\sm}$ \\
\noalign{\medskip}
\hline
\noalign{\smallskip}
$I_{1s}$ & $\frac{1}{2}(H_+^2 + H_-^2)(m_{\ell}^2 + 3 q^2)$ \\
\noalign{\medskip}
$I_{1c}$ & $4 m_{\ell}^2 H_t^2 + 2 H_0^2 (m_{\ell}^2 + q^2)$ \\
\noalign{\medskip}
$I_{2s}$ & $- \frac{1}{2}(H_+^2 + H_-^2)(m_{\ell}^2 - q^2)$ \\
\noalign{\medskip}
$I_{2c}$ & $2 H_0^2 (m_{\ell}^2 - q^2)$ \\
\noalign{\medskip}
$I_{3}$ & $2 H_+ H_- (m_{\ell}^2 - q^2)$ \\
\noalign{\medskip}
$I_{4}$ & $H_0 (H_+ + H_-) (m_{\ell}^2 - q^2)$ \\
\noalign{\medskip}
$I_{5}$ & $-2 H_t (H_+ + H_-) m_{\ell}^2 - 2 H_0 (H_+ - H_-) q^2$ \\
\noalign{\medskip}
$I_{6s}$ & $2 (H_+^2 - H_-^2) q^2$ \\
\noalign{\medskip}
$I_{6c}$ & $- 8 H_t H_0 m_{\ell}^2$ \\
\noalign{\medskip}
$I_{7}$ & $0$ \\
\noalign{\medskip}
$I_{8}$ & $0$ \\
\noalign{\medskip}
$I_{9}$ & $0$ \\
\noalign{\medskip}
\hline
\hline
\end{tabular}
\end{table}
%
%
\begin{table}[htp]
\caption{\small Angular coefficient functions for   $\bar B \to V (P_1 P_2) \ell^- \bar \nu_\ell$:  NP term with the R operator, and interference  SM-NP with R operator. }\label{tab:R}
\vspace{0.3cm}
\centering
\begin{tabular}{ccc}
\hline
\hline
\noalign{\medskip}
$i$ & $I_i^{\np,R}$ & $I_i^{\inter,R}$  \\
\noalign{\medskip}
\hline
\noalign{\smallskip}
$I_{1s}$ & $\frac{1}{2}(H_+^2 + H_-^2)(m_{\ell}^2 + 3 q^2)$ & $-H_- H_+ (m_\ell^2 + 3 q^2)$\\
\noalign{\medskip}
$I_{1c}$ & $4 m_{\ell}^2 H_t^2 + 2 H_0^2 (m_{\ell}^2 + q^2)$& $-2 (2 H_t^2 m_\ell^2 + H_0^2 (m_\ell^2 + q^2))$\\
\noalign{\medskip}
$I_{2s}$ & $- \frac{1}{2}(H_+^2 + H_-^2)(m_{\ell}^2 - q^2)$ & $H_- H_+ (m_{\ell}^2 - q^2)$ \\
\noalign{\medskip}
$I_{2c}$ & $2 H_0^2 (m_{\ell}^2 - q^2)$ & $-2H_0^2 (m_{\ell}^2 - q^2)$ \\
\noalign{\medskip}
$I_{3}$ & $2 H_+ H_- (m_{\ell}^2 - q^2)$ & $-(H_+^2 + H_-^2) (m_{\ell}^2 - q^2)$\\
\noalign{\medskip}
$I_{4}$ & $H_0 (H_+ + H_-) (m_{\ell}^2 - q^2)$ & $-H_0(H_+ + H_-) (m_{\ell}^2 - q^2)$ \\
\noalign{\medskip}
$I_{5}$ & $-2 H_t (H_+ + H_-) m_{\ell}^2 + 2 H_0 (H_+ - H_-) q^2$ & $2 H_t (H_+ + H_-) m_{\ell}^2$\\
\noalign{\medskip}
$I_{6s}$ & $- 2 (H_+^2 - H_-^2) q^2$ &  $0$\\
\noalign{\medskip}
$I_{6c}$ & $- 8 H_t H_0 m_{\ell}^2$ & $8H_0 H_t m_\ell^2$\\
\noalign{\medskip}
$I_{7}$ & $0$ & $2(H_+-H_-)H_t m_\ell^2$\\
\noalign{\medskip}
$I_{8}$ & $0$ & $-H_0(H_+-H_-)(m_{\ell}^2 - q^2)$\\
\noalign{\medskip}
$I_{9}$ & $0$ & $-(H_+^2 - H_-^2) (m_{\ell}^2 - q^2)$\\
\noalign{\medskip}
\hline
\hline
\end{tabular}
\end{table}
\begin{table}[htp]
\caption{\small Angular coefficient functions for   $\bar B \to V (P_1 P_2) \ell^- \bar \nu_\ell$:  NP term with the P operator, and interference  SM-NP with P operator. }\label{tab:P}
\vspace{0.3cm}
\centering
\begin{tabular}{ccc}
\hline
\hline
\noalign{\medskip}
$i$ & $I_i^{\np,P}$ & $I_i^{\inter,P}$  \\
\noalign{\medskip}
\hline
\noalign{\medskip}
$I_{1s}$ & $0$ & $0$ \\
\noalign{\medskip}
$I_{1c}$ & $4 H_t^2 \frac{q^4}{(m_b + m_U)^2}$ & $4 H_t^2 \frac{m_{\ell} q^2}{m_b + m_U}$ \\
\noalign{\medskip}
$I_{2s}$ & $0$ & $0$  \\
\noalign{\medskip}
$I_{2c}$ & $0$ & $0$  \\
\noalign{\medskip}
$I_{3}$ & $0$ & $0$  \\
\noalign{\medskip}
$I_{4}$ & $0$ & $0$ \\
\noalign{\medskip}
$I_{5}$ & $0$ & $-H_t(H_+ + H_-) \frac{m_{\ell} q^2}{m_b + m_U}$  \\
\noalign{\medskip}
$I_{6s}$ & $0$ & $0$ \\
\noalign{\medskip}
$I_{6c}$ & $0$ & $-4 H_t H_0 \frac{m_{\ell} q^2}{m_b + m_U}$ \\
\noalign{\medskip}
$I_{7}$ & $0$ & $- H_t (H_+ - H_-) \frac{m_{\ell} q^2}{m_b + m_U}$  \\
\noalign{\medskip}
$I_{8}$ & 0 & 0 \\
\noalign{\medskip}
$I_{9}$ & 0 & 0 \\
\noalign{\medskip}
\hline
\hline
\end{tabular}
\end{table}
\begin{table}[htp]
\caption{ \small Angular coefficient functions   for  $\bar B \to V (P_1 P_2)   \ell^- \bar \nu_\ell$: NP term with the T operator  and  interference  SM-NP with T operator.}\label{tab:T}
\vspace{0.3cm}
\centering
\begin{tabular}{ccc}
\hline
\hline
\noalign{\medskip}
$i$ & $I_i^{\np,T}$ & $I_i^{\inter, T}$ \\
\noalign{\medskip}
\hline
\noalign{\bigskip}
$I_{1s}$ & $2[(H_+^\np)^2+(H_-^\np)^2](3 m_\ell^2 + q^2)$ & $-4 ( H_+^\np H_+ + H_-^\np H_- ) m_\ell \sqrt{q^2}$ \\
\noalign{\medskip}
$I_{1c}$ & $\frac{1}{8} (H_L^\np)^2 (m_\ell^2 + q^2)$ & $- H_L^\np H_0 m_\ell \sqrt{q^2}$ \\
\noalign{\medskip}
$I_{2s}$ & $2[(H_+^\np)^2+(H_-^\np)^2](m_\ell^2 - q^2)$ & $0$ \\
\noalign{\medskip}
$I_{2c}$ & $\frac{1}{8} (H_L^\np)^2 (q^2 - m_\ell^2)$ & $0$ \\
\noalign{\medskip}
$I_{3}$ & $8 H_+^\np H_-^\np (q^2 - m_\ell^2)$ & $0$ \\
\noalign{\medskip}
$I_{4}$ & $\frac{1}{2} H_L^\np (H_+^\np + H_-^\np) (q^2 - m_\ell^2)$ & $0$ \\
\noalign{\medskip}
$I_{5}$ & $- H_L^\np (H_+^\np - H_-^\np) m_\ell^2$ & $\frac{1}{4} [H_L^\np (H_+ - H_-) + 8 H_+^\np (H_t + H_0) + 8 H_-^\np(H_t - H_0)] m_{\ell} \sqrt{q^2} $ \\
\noalign{\medskip}
$I_{6s}$ & $8 [(H_+^\np)^2 - (H_-^\np)^2] m_\ell^2$ & $-4 ( H_+^\np H_+ - H_-^\np H_- ) m_\ell \sqrt{q^2}$ \\
\noalign{\medskip}
$I_{6c}$ & $0$ & $H_L^\np H_t m_{\ell} \sqrt{q^2}$ \\
\noalign{\medskip}
$I_{7}$ & $0$ & $\frac{1}{4} [H_L^\np (H_+ + H_-) - 8 H_+^\np (H_t + H_0) + 8 H_-^\np(H_t - H_0)] m_{\ell} \sqrt{q^2} $ \\
\noalign{\medskip}
$I_{8}$ &  0& 0 \\
\noalign{\medskip}
$I_{9}$ &  0& 0 \\
\noalign{\medskip}
\hline
\hline
\end{tabular}
\end{table}
\begin{table}[htp]
\caption{\small Angular coefficient functions for   $\bar B \to V (P_1 P_2) \ell^- \bar \nu_\ell$:  P-R, R-T and P-T  interferences. }\label{tab:intvari}
\vspace{0.3cm}
\centering
\begin{tabular}{cccc}
\hline
\hline
\noalign{\medskip}
$i$ & $I_i^{\inter,PR}$ & $I_i^{\inter, RT}$ & $I_i^{\inter, PT}$ \\
\noalign{\medskip}
\hline
\noalign{\medskip}
$I_{1s}$ & $0$  &$4(H_-^{\np} H_+ + H_- H_+^{\np}) m_\ell \sqrt{q^2}$  & $0$\\
\noalign{\medskip}
$I_{1c}$ & $-4H_t^2 m_\ell \frac{ q^2}{m_b + m_U}$& $H_0\, H_L^{\np}m_\ell \sqrt{q^2}$ & $0$\\
\noalign{\medskip}
$I_{2s}$ & $0$ & $0$ & $0$ \\
\noalign{\medskip}
$I_{2c}$ &$0$   & $0$  & $0$\\
\noalign{\medskip}
$I_{3}$ &  $0$ & $0$ & $0$\\
\noalign{\medskip}
$I_{4}$ & $0$ & $0$ & $0$ \\
\noalign{\medskip}
$I_{5}$ & $(H_++H_-)H_t  m_\ell \frac{ q^2}{m_b + m_U}$ &  \,\,$\frac{1}{4} [H_L^\np (H_+ - H_-) - 8 H_+^\np (H_t + H_0) - 8 H_-^\np(H_t - H_0)] m_{\ell} \sqrt{q^2} $ \,\,
&\,\,  $ 2H_t(H_+^{\np} + H_-^{\np}) \frac{ (q^2)^{3/2}}{m_b + m_U}$   \\
\noalign{\medskip}
$I_{6s}$ &  $0$ & $4(-H_-^{\np} H_+ + H_- H_+^{\np}) m_\ell \sqrt{q^2}$ & $0$\\
\noalign{\medskip}
$I_{6c}$ & $4H_0 H_t m_\ell \frac{ q^2}{m_b + m_U}$ & $-H_t\, H_L^{\np}m_\ell \sqrt{q^2}$
 & \,\,$ H_t\, H_L^{\np} \frac{ (q^2)^{3/2}}{m_b + m_U}$\\
\noalign{\medskip}
$I_{7}$ & $-H_t (H_+ -H_-) m_\ell \frac{ q^2}{m_b + m_U}$ & $\frac{1}{4} [H_L^\np (H_+ + H_-) - 8 H_+^\np (H_t + H_0) + 8 H_-^\np(H_t - H_0)] m_{\ell} \sqrt{q^2} $ & \,\, $ 2H_t(H_+^{\np} - H_-^{\np}) \frac{ (q^2)^{3/2}}{m_b + m_U}$ \\
\noalign{\medskip}
$I_{8}$ &$0$  & $0$& $0$\\
\noalign{\medskip}
$I_{9}$ & $0$ &$0$  &  $0$\\
\noalign{\medskip}
\hline
\hline
\end{tabular}
\end{table}
\end{widetext}

\bibliographystyle{apsrev4-1}
\bibliography{refFFNP}

\begin{thebibliography}{27}%
\makeatletter
\providecommand \@ifxundefined [1]{%
 \@ifx{#1\undefined}
}%
\providecommand \@ifnum [1]{%
 \ifnum #1\expandafter \@firstoftwo
 \else \expandafter \@secondoftwo
 \fi
}%
\providecommand \@ifx [1]{%
 \ifx #1\expandafter \@firstoftwo
 \else \expandafter \@secondoftwo
 \fi
}%
\providecommand \natexlab [1]{#1}%
\providecommand \enquote  [1]{``#1''}%
\providecommand \bibnamefont  [1]{#1}%
\providecommand \bibfnamefont [1]{#1}%
\providecommand \citenamefont [1]{#1}%
\providecommand \href@noop [0]{\@secondoftwo}%
\providecommand \href [0]{\begingroup \@sanitize@url \@href}%
\providecommand \@href[1]{\@@startlink{#1}\@@href}%
\providecommand \@@href[1]{\endgroup#1\@@endlink}%
\providecommand \@sanitize@url [0]{\catcode `\\12\catcode `\$12\catcode
  `\&12\catcode `\#12\catcode `\^12\catcode `\_12\catcode `\%12\relax}%
\providecommand \@@startlink[1]{}%
\providecommand \@@endlink[0]{}%
\providecommand \url  [0]{\begingroup\@sanitize@url \@url }%
\providecommand \@url [1]{\endgroup\@href {#1}{\urlprefix }}%
\providecommand \urlprefix  [0]{URL }%
\providecommand \Eprint [0]{\href }%
\providecommand \doibase [0]{http://dx.doi.org/}%
\providecommand \selectlanguage [0]{\@gobble}%
\providecommand \bibinfo  [0]{\@secondoftwo}%
\providecommand \bibfield  [0]{\@secondoftwo}%
\providecommand \translation [1]{[#1]}%
\providecommand \BibitemOpen [0]{}%
\providecommand \bibitemStop [0]{}%
\providecommand \bibitemNoStop [0]{.\EOS\space}%
\providecommand \EOS [0]{\spacefactor3000\relax}%
\providecommand \BibitemShut  [1]{\csname bibitem#1\endcsname}%
\let\auto@bib@innerbib\@empty
\bibitem [{\citenamefont {Prim}\ \emph {et~al.}(2023)\citenamefont {Prim} \emph
  {et~al.}}]{Belle:2023xgj}%
  \BibitemOpen
  \bibfield  {author} {\bibinfo {author} {\bibfnamefont {M.~T.}\ \bibnamefont
  {Prim}} \emph {et~al.} (\bibinfo {collaboration} {Belle}),\ }\href@noop {} {\
   (\bibinfo {year} {2023})},\ \Eprint {http://arxiv.org/abs/2310.20286}
  {arXiv:2310.20286 [hep-ex]} \BibitemShut {NoStop}%
\bibitem [{\citenamefont {Amhis}\ \emph {et~al.}(2023)\citenamefont {Amhis}
  \emph {et~al.}}]{HFLAV:2022pwe}%
  \BibitemOpen
  \bibfield  {author} {\bibinfo {author} {\bibfnamefont {Y.}~\bibnamefont
  {Amhis}} \emph {et~al.},\ }\href {\doibase 10.1103/PhysRevD.107.052008}
  {\bibfield  {journal} {\bibinfo  {journal} {Phys. Rev. D}\ }\textbf {\bibinfo
  {volume} {107}},\ \bibinfo {pages} {052008} (\bibinfo {year} {2023})},\
  \Eprint {http://arxiv.org/abs/2206.07501} {arXiv:2206.07501 [hep-ex]}
  \BibitemShut {NoStop}%
\bibitem [{\citenamefont {De~Fazio}(2023)}]{DeFazio:2023lmy}%
  \BibitemOpen
  \bibfield  {author} {\bibinfo {author} {\bibfnamefont {F.}~\bibnamefont
  {De~Fazio}}\ }(\bibinfo {year} {2023})\ \Eprint
  {http://arxiv.org/abs/2311.02987} {arXiv:2311.02987 [hep-ph]} \BibitemShut
  {NoStop}%
\bibitem [{\citenamefont {Buras}(2020)}]{Buras:2020xsm}%
  \BibitemOpen
  \bibfield  {author} {\bibinfo {author} {\bibfnamefont {A.}~\bibnamefont
  {Buras}},\ }\href {\doibase 10.1017/9781139524100} {\emph {\bibinfo {title}
  {{Gauge Theory of Weak Decays}}}}\ (\bibinfo  {publisher} {Cambridge
  University Press},\ \bibinfo {year} {2020})\BibitemShut {NoStop}%
\bibitem [{\citenamefont {Lees}\ \emph {et~al.}(2012)\citenamefont {Lees} \emph
  {et~al.}}]{Lees:2012xj}%
  \BibitemOpen
  \bibfield  {author} {\bibinfo {author} {\bibfnamefont {J.~P.}\ \bibnamefont
  {Lees}} \emph {et~al.} (\bibinfo {collaboration} {BaBar}),\ }\href {\doibase
  10.1103/PhysRevLett.109.101802} {\bibfield  {journal} {\bibinfo  {journal}
  {Phys. Rev. Lett.}\ }\textbf {\bibinfo {volume} {109}},\ \bibinfo {pages}
  {101802} (\bibinfo {year} {2012})},\ \Eprint {http://arxiv.org/abs/1205.5442}
  {arXiv:1205.5442 [hep-ex]} \BibitemShut {NoStop}%
\bibitem [{\citenamefont {Lees}\ \emph {et~al.}(2013)\citenamefont {Lees} \emph
  {et~al.}}]{BaBar:2013mob}%
  \BibitemOpen
  \bibfield  {author} {\bibinfo {author} {\bibfnamefont {J.~P.}\ \bibnamefont
  {Lees}} \emph {et~al.} (\bibinfo {collaboration} {BaBar}),\ }\href {\doibase
  10.1103/PhysRevD.88.072012} {\bibfield  {journal} {\bibinfo  {journal} {Phys.
  Rev. D}\ }\textbf {\bibinfo {volume} {88}},\ \bibinfo {pages} {072012}
  (\bibinfo {year} {2013})},\ \Eprint {http://arxiv.org/abs/1303.0571}
  {arXiv:1303.0571 [hep-ex]} \BibitemShut {NoStop}%
\bibitem [{\citenamefont {Huschle}\ \emph {et~al.}(2015)\citenamefont {Huschle}
  \emph {et~al.}}]{Belle:2015qfa}%
  \BibitemOpen
  \bibfield  {author} {\bibinfo {author} {\bibfnamefont {M.}~\bibnamefont
  {Huschle}} \emph {et~al.} (\bibinfo {collaboration} {Belle}),\ }\href
  {\doibase 10.1103/PhysRevD.92.072014} {\bibfield  {journal} {\bibinfo
  {journal} {Phys. Rev. D}\ }\textbf {\bibinfo {volume} {92}},\ \bibinfo
  {pages} {072014} (\bibinfo {year} {2015})},\ \Eprint
  {http://arxiv.org/abs/1507.03233} {arXiv:1507.03233 [hep-ex]} \BibitemShut
  {NoStop}%
\bibitem [{\citenamefont {Aaij}\ \emph {et~al.}(2015)\citenamefont {Aaij} \emph
  {et~al.}}]{LHCb:2015gmp}%
  \BibitemOpen
  \bibfield  {author} {\bibinfo {author} {\bibfnamefont {R.}~\bibnamefont
  {Aaij}} \emph {et~al.} (\bibinfo {collaboration} {LHCb}),\ }\href {\doibase
  10.1103/PhysRevLett.115.111803} {\bibfield  {journal} {\bibinfo  {journal}
  {Phys. Rev. Lett.}\ }\textbf {\bibinfo {volume} {115}},\ \bibinfo {pages}
  {111803} (\bibinfo {year} {2015})},\ \bibinfo {note} {[Erratum:
  Phys.Rev.Lett. 115, 159901 (2015)]},\ \Eprint
  {http://arxiv.org/abs/1506.08614} {arXiv:1506.08614 [hep-ex]} \BibitemShut
  {NoStop}%
\bibitem [{\citenamefont {Hirose}\ \emph {et~al.}(2017)\citenamefont {Hirose}
  \emph {et~al.}}]{Belle:2016dyj}%
  \BibitemOpen
  \bibfield  {author} {\bibinfo {author} {\bibfnamefont {S.}~\bibnamefont
  {Hirose}} \emph {et~al.} (\bibinfo {collaboration} {Belle}),\ }\href
  {\doibase 10.1103/PhysRevLett.118.211801} {\bibfield  {journal} {\bibinfo
  {journal} {Phys. Rev. Lett.}\ }\textbf {\bibinfo {volume} {118}},\ \bibinfo
  {pages} {211801} (\bibinfo {year} {2017})},\ \Eprint
  {http://arxiv.org/abs/1612.00529} {arXiv:1612.00529 [hep-ex]} \BibitemShut
  {NoStop}%
\bibitem [{\citenamefont {Aaij}\ \emph
  {et~al.}(2018{\natexlab{a}})\citenamefont {Aaij} \emph
  {et~al.}}]{LHCb:2017smo}%
  \BibitemOpen
  \bibfield  {author} {\bibinfo {author} {\bibfnamefont {R.}~\bibnamefont
  {Aaij}} \emph {et~al.} (\bibinfo {collaboration} {LHCb}),\ }\href {\doibase
  10.1103/PhysRevLett.120.171802} {\bibfield  {journal} {\bibinfo  {journal}
  {Phys. Rev. Lett.}\ }\textbf {\bibinfo {volume} {120}},\ \bibinfo {pages}
  {171802} (\bibinfo {year} {2018}{\natexlab{a}})},\ \Eprint
  {http://arxiv.org/abs/1708.08856} {arXiv:1708.08856 [hep-ex]} \BibitemShut
  {NoStop}%
\bibitem [{\citenamefont {Aaij}\ \emph
  {et~al.}(2018{\natexlab{b}})\citenamefont {Aaij} \emph
  {et~al.}}]{LHCb:2017rln}%
  \BibitemOpen
  \bibfield  {author} {\bibinfo {author} {\bibfnamefont {R.}~\bibnamefont
  {Aaij}} \emph {et~al.} (\bibinfo {collaboration} {LHCb}),\ }\href {\doibase
  10.1103/PhysRevD.97.072013} {\bibfield  {journal} {\bibinfo  {journal} {Phys.
  Rev. D}\ }\textbf {\bibinfo {volume} {97}},\ \bibinfo {pages} {072013}
  (\bibinfo {year} {2018}{\natexlab{b}})},\ \Eprint
  {http://arxiv.org/abs/1711.02505} {arXiv:1711.02505 [hep-ex]} \BibitemShut
  {NoStop}%
\bibitem [{\citenamefont {Caria}\ \emph {et~al.}(2020)\citenamefont {Caria}
  \emph {et~al.}}]{Belle:2019rba}%
  \BibitemOpen
  \bibfield  {author} {\bibinfo {author} {\bibfnamefont {G.}~\bibnamefont
  {Caria}} \emph {et~al.} (\bibinfo {collaboration} {Belle}),\ }\href {\doibase
  10.1103/PhysRevLett.124.161803} {\bibfield  {journal} {\bibinfo  {journal}
  {Phys. Rev. Lett.}\ }\textbf {\bibinfo {volume} {124}},\ \bibinfo {pages}
  {161803} (\bibinfo {year} {2020})},\ \Eprint
  {http://arxiv.org/abs/1910.05864} {arXiv:1910.05864 [hep-ex]} \BibitemShut
  {NoStop}%
\bibitem [{\citenamefont {Gambino}\ \emph {et~al.}(2020)\citenamefont {Gambino}
  \emph {et~al.}}]{Gambino:2020jvv}%
  \BibitemOpen
  \bibfield  {author} {\bibinfo {author} {\bibfnamefont {P.}~\bibnamefont
  {Gambino}} \emph {et~al.},\ }\href {\doibase 10.1140/epjc/s10052-020-08490-x}
  {\bibfield  {journal} {\bibinfo  {journal} {Eur. Phys. J. C}\ }\textbf
  {\bibinfo {volume} {80}},\ \bibinfo {pages} {966} (\bibinfo {year} {2020})},\
  \Eprint {http://arxiv.org/abs/2006.07287} {arXiv:2006.07287 [hep-ph]}
  \BibitemShut {NoStop}%
\bibitem [{\citenamefont {Colangelo}\ and\ \citenamefont
  {De~Fazio}(2017)}]{Colangelo:2016ymy}%
  \BibitemOpen
  \bibfield  {author} {\bibinfo {author} {\bibfnamefont {P.}~\bibnamefont
  {Colangelo}}\ and\ \bibinfo {author} {\bibfnamefont {F.}~\bibnamefont
  {De~Fazio}},\ }\href {\doibase 10.1103/PhysRevD.95.011701} {\bibfield
  {journal} {\bibinfo  {journal} {Phys. Rev.}\ }\textbf {\bibinfo {volume}
  {D95}},\ \bibinfo {pages} {011701} (\bibinfo {year} {2017})},\ \Eprint
  {http://arxiv.org/abs/1611.07387} {arXiv:1611.07387 [hep-ph]} \BibitemShut
  {NoStop}%
\bibitem [{\citenamefont {Colangelo}\ and\ \citenamefont
  {De~Fazio}(2018)}]{Colangelo:2018cnj}%
  \BibitemOpen
  \bibfield  {author} {\bibinfo {author} {\bibfnamefont {P.}~\bibnamefont
  {Colangelo}}\ and\ \bibinfo {author} {\bibfnamefont {F.}~\bibnamefont
  {De~Fazio}},\ }\href {\doibase 10.1007/JHEP06(2018)082} {\bibfield  {journal}
  {\bibinfo  {journal} {JHEP}\ }\textbf {\bibinfo {volume} {06}},\ \bibinfo
  {pages} {082} (\bibinfo {year} {2018})},\ \Eprint
  {http://arxiv.org/abs/1801.10468} {arXiv:1801.10468 [hep-ph]} \BibitemShut
  {NoStop}%
\bibitem [{\citenamefont {Bhattacharya}\ \emph {et~al.}(2019)\citenamefont
  {Bhattacharya}, \citenamefont {Nandi},\ and\ \citenamefont
  {Kumar~Patra}}]{Bhattacharya:2018kig}%
  \BibitemOpen
  \bibfield  {author} {\bibinfo {author} {\bibfnamefont {S.}~\bibnamefont
  {Bhattacharya}}, \bibinfo {author} {\bibfnamefont {S.}~\bibnamefont {Nandi}},
  \ and\ \bibinfo {author} {\bibfnamefont {S.}~\bibnamefont {Kumar~Patra}},\
  }\href {\doibase 10.1140/epjc/s10052-019-6767-7} {\bibfield  {journal}
  {\bibinfo  {journal} {Eur. Phys. J. C}\ }\textbf {\bibinfo {volume} {79}},\
  \bibinfo {pages} {268} (\bibinfo {year} {2019})},\ \Eprint
  {http://arxiv.org/abs/1805.08222} {arXiv:1805.08222 [hep-ph]} \BibitemShut
  {NoStop}%
\bibitem [{\citenamefont {Murgui}\ \emph {et~al.}(2019)\citenamefont {Murgui},
  \citenamefont {Pe\~nuelas}, \citenamefont {Jung},\ and\ \citenamefont
  {Pich}}]{Murgui:2019czp}%
  \BibitemOpen
  \bibfield  {author} {\bibinfo {author} {\bibfnamefont {C.}~\bibnamefont
  {Murgui}}, \bibinfo {author} {\bibfnamefont {A.}~\bibnamefont {Pe\~nuelas}},
  \bibinfo {author} {\bibfnamefont {M.}~\bibnamefont {Jung}}, \ and\ \bibinfo
  {author} {\bibfnamefont {A.}~\bibnamefont {Pich}},\ }\href {\doibase
  10.1007/JHEP09(2019)103} {\bibfield  {journal} {\bibinfo  {journal} {JHEP}\
  }\textbf {\bibinfo {volume} {09}},\ \bibinfo {pages} {103} (\bibinfo {year}
  {2019})},\ \Eprint {http://arxiv.org/abs/1904.09311} {arXiv:1904.09311
  [hep-ph]} \BibitemShut {NoStop}%
\bibitem [{\citenamefont {Be\v{c}irevi\'c}\ \emph {et~al.}(2019)\citenamefont
  {Be\v{c}irevi\'c}, \citenamefont {Fedele}, \citenamefont
  {Ni\v{s}and\v{z}i\'c},\ and\ \citenamefont {Tayduganov}}]{Becirevic:2019tpx}%
  \BibitemOpen
  \bibfield  {author} {\bibinfo {author} {\bibfnamefont {D.}~\bibnamefont
  {Be\v{c}irevi\'c}}, \bibinfo {author} {\bibfnamefont {M.}~\bibnamefont
  {Fedele}}, \bibinfo {author} {\bibfnamefont {I.}~\bibnamefont
  {Ni\v{s}and\v{z}i\'c}}, \ and\ \bibinfo {author} {\bibfnamefont
  {A.}~\bibnamefont {Tayduganov}},\ }\href@noop {} {\  (\bibinfo {year}
  {2019})},\ \Eprint {http://arxiv.org/abs/1907.02257} {arXiv:1907.02257
  [hep-ph]} \BibitemShut {NoStop}%
\bibitem [{\citenamefont {Bobeth}\ \emph {et~al.}(2021)\citenamefont {Bobeth},
  \citenamefont {Bordone}, \citenamefont {Gubernari}, \citenamefont {Jung},\
  and\ \citenamefont {van Dyk}}]{Bobeth:2021lya}%
  \BibitemOpen
  \bibfield  {author} {\bibinfo {author} {\bibfnamefont {C.}~\bibnamefont
  {Bobeth}}, \bibinfo {author} {\bibfnamefont {M.}~\bibnamefont {Bordone}},
  \bibinfo {author} {\bibfnamefont {N.}~\bibnamefont {Gubernari}}, \bibinfo
  {author} {\bibfnamefont {M.}~\bibnamefont {Jung}}, \ and\ \bibinfo {author}
  {\bibfnamefont {D.}~\bibnamefont {van Dyk}},\ }\href {\doibase
  10.1140/epjc/s10052-021-09724-2} {\bibfield  {journal} {\bibinfo  {journal}
  {Eur. Phys. J. C}\ }\textbf {\bibinfo {volume} {81}},\ \bibinfo {pages} {984}
  (\bibinfo {year} {2021})},\ \Eprint {http://arxiv.org/abs/2104.02094}
  {arXiv:2104.02094 [hep-ph]} \BibitemShut {NoStop}%
\bibitem [{\citenamefont {Colangelo}\ \emph {et~al.}(2019)\citenamefont
  {Colangelo}, \citenamefont {De~Fazio},\ and\ \citenamefont
  {Loparco}}]{Colangelo:2019axi}%
  \BibitemOpen
  \bibfield  {author} {\bibinfo {author} {\bibfnamefont {P.}~\bibnamefont
  {Colangelo}}, \bibinfo {author} {\bibfnamefont {F.}~\bibnamefont {De~Fazio}},
  \ and\ \bibinfo {author} {\bibfnamefont {F.}~\bibnamefont {Loparco}},\ }\href
  {\doibase 10.1103/PhysRevD.100.075037} {\bibfield  {journal} {\bibinfo
  {journal} {Phys. Rev. D}\ }\textbf {\bibinfo {volume} {100}},\ \bibinfo
  {pages} {075037} (\bibinfo {year} {2019})},\ \Eprint
  {http://arxiv.org/abs/1906.07068} {arXiv:1906.07068 [hep-ph]} \BibitemShut
  {NoStop}%
\bibitem [{\citenamefont {Colangelo}\ \emph {et~al.}(2021)\citenamefont
  {Colangelo}, \citenamefont {De~Fazio},\ and\ \citenamefont
  {Loparco}}]{Colangelo:2021dnv}%
  \BibitemOpen
  \bibfield  {author} {\bibinfo {author} {\bibfnamefont {P.}~\bibnamefont
  {Colangelo}}, \bibinfo {author} {\bibfnamefont {F.}~\bibnamefont {De~Fazio}},
  \ and\ \bibinfo {author} {\bibfnamefont {F.}~\bibnamefont {Loparco}},\ }\href
  {\doibase 10.1103/PhysRevD.103.075019} {\bibfield  {journal} {\bibinfo
  {journal} {Phys. Rev. D}\ }\textbf {\bibinfo {volume} {103}},\ \bibinfo
  {pages} {075019} (\bibinfo {year} {2021})},\ \Eprint
  {http://arxiv.org/abs/2102.05365} {arXiv:2102.05365 [hep-ph]} \BibitemShut
  {NoStop}%
\bibitem [{\citenamefont {Buchmuller}\ and\ \citenamefont
  {Wyler}(1986)}]{Buchmuller:1985jz}%
  \BibitemOpen
  \bibfield  {author} {\bibinfo {author} {\bibfnamefont {W.}~\bibnamefont
  {Buchmuller}}\ and\ \bibinfo {author} {\bibfnamefont {D.}~\bibnamefont
  {Wyler}},\ }\href {\doibase 10.1016/0550-3213(86)90262-2} {\bibfield
  {journal} {\bibinfo  {journal} {Nucl. Phys.}\ }\textbf {\bibinfo {volume}
  {B268}},\ \bibinfo {pages} {621} (\bibinfo {year} {1986})}\BibitemShut
  {NoStop}%
\bibitem [{\citenamefont {Grzadkowski}\ \emph {et~al.}(2010)\citenamefont
  {Grzadkowski}, \citenamefont {Iskrzynski}, \citenamefont {Misiak},\ and\
  \citenamefont {Rosiek}}]{Grzadkowski:2010es}%
  \BibitemOpen
  \bibfield  {author} {\bibinfo {author} {\bibfnamefont {B.}~\bibnamefont
  {Grzadkowski}}, \bibinfo {author} {\bibfnamefont {M.}~\bibnamefont
  {Iskrzynski}}, \bibinfo {author} {\bibfnamefont {M.}~\bibnamefont {Misiak}},
  \ and\ \bibinfo {author} {\bibfnamefont {J.}~\bibnamefont {Rosiek}},\ }\href
  {\doibase 10.1007/JHEP10(2010)085} {\bibfield  {journal} {\bibinfo  {journal}
  {JHEP}\ }\textbf {\bibinfo {volume} {10}},\ \bibinfo {pages} {085} (\bibinfo
  {year} {2010})},\ \Eprint {http://arxiv.org/abs/1008.4884} {arXiv:1008.4884
  [hep-ph]} \BibitemShut {NoStop}%
\bibitem [{\citenamefont {Workman}\ and\ \citenamefont
  {Others}(2022)}]{Workman:2022ynf}%
  \BibitemOpen
  \bibfield  {author} {\bibinfo {author} {\bibfnamefont {R.~L.}\ \bibnamefont
  {Workman}}\ and\ \bibinfo {author} {\bibnamefont {Others}} (\bibinfo
  {collaboration} {Particle Data Group}),\ }\href {\doibase
  10.1093/ptep/ptac097} {\bibfield  {journal} {\bibinfo  {journal} {PTEP}\
  }\textbf {\bibinfo {volume} {2022}},\ \bibinfo {pages} {083C01} (\bibinfo
  {year} {2022})}\BibitemShut {NoStop}%
\bibitem [{\citenamefont {Caprini}\ \emph {et~al.}(1998)\citenamefont
  {Caprini}, \citenamefont {Lellouch},\ and\ \citenamefont
  {Neubert}}]{Caprini:1997mu}%
  \BibitemOpen
  \bibfield  {author} {\bibinfo {author} {\bibfnamefont {I.}~\bibnamefont
  {Caprini}}, \bibinfo {author} {\bibfnamefont {L.}~\bibnamefont {Lellouch}}, \
  and\ \bibinfo {author} {\bibfnamefont {M.}~\bibnamefont {Neubert}},\ }\href
  {\doibase 10.1016/S0550-3213(98)00350-2} {\bibfield  {journal} {\bibinfo
  {journal} {Nucl. Phys. B}\ }\textbf {\bibinfo {volume} {530}},\ \bibinfo
  {pages} {153} (\bibinfo {year} {1998})},\ \Eprint
  {http://arxiv.org/abs/hep-ph/9712417} {arXiv:hep-ph/9712417} \BibitemShut
  {NoStop}%
\bibitem [{\citenamefont {Abdesselam}\ \emph {et~al.}(2017)\citenamefont
  {Abdesselam} \emph {et~al.}}]{Belle:2017rcc}%
  \BibitemOpen
  \bibfield  {author} {\bibinfo {author} {\bibfnamefont {A.}~\bibnamefont
  {Abdesselam}} \emph {et~al.} (\bibinfo {collaboration} {Belle}),\ }\href@noop
  {} {\  (\bibinfo {year} {2017})},\ \Eprint {http://arxiv.org/abs/1702.01521}
  {arXiv:1702.01521 [hep-ex]} \BibitemShut {NoStop}%
\bibitem [{\citenamefont {Kapoor}\ \emph {et~al.}(2024)\citenamefont {Kapoor},
  \citenamefont {Huang},\ and\ \citenamefont {Kou}}]{Kapoor:2024ufg}%
  \BibitemOpen
  \bibfield  {author} {\bibinfo {author} {\bibfnamefont {T.}~\bibnamefont
  {Kapoor}}, \bibinfo {author} {\bibfnamefont {Z.-R.}\ \bibnamefont {Huang}}, \
  and\ \bibinfo {author} {\bibfnamefont {E.}~\bibnamefont {Kou}},\ }\href@noop
  {} {\  (\bibinfo {year} {2024})},\ \Eprint {http://arxiv.org/abs/2401.11636}
  {arXiv:2401.11636 [hep-ph]} \BibitemShut {NoStop}%
\end{thebibliography}%
\end{document}